\documentclass[twocolumn,prX]{revtex4}
\usepackage{multirow}
\usepackage{graphicx}
\usepackage{amsmath}
\usepackage{enumerate}
\usepackage{epstopdf}
\usepackage{times}
\usepackage{color}

\usepackage[colorlinks,bookmarks=false,citecolor=blue,linkcolor=red,urlcolor=blue]{hyperref}

\begin{document}
\title{Detecting non-magnetic excitations in quantum magnets}
\author{Zhihao Hao}
\affiliation{Department of Physics and Astronomy, Johns Hopkins
University, Baltimore, Maryland 21218, USA}
\affiliation{Department of Physics and Astronomy, University of Waterloo, Waterloo, Ontario N2L 3G1, Canada}
\begin{abstract}
Many unconventional quantum phases host special non-magnetic excitations such as photons and visons. We discuss two possible ways to detect these excitations experimentally. Firstly, spin-lattice coupling mixes the excitations with phonons. The phonon spectral function acquires new features that can be detected by neutron or X-ray scattering. Secondly, valence-bond fluctuations translate into charge density fluctuations on non-bipartite lattices. Such charge fluctuations can be characterized by conventional spectroscopies such as Terahertz spectroscopy. Observation of exotic singlet excitations would provide positive identification of unconventional quantum phases in frustrated antiferromagnets.
\end{abstract}

\maketitle

The search for exotic quantum phases \cite{LMM} in frustrated antiferromagnets has been one of the main challenges in the field of strongly correlated systems. Such phases are believed to emerge when long-range order is destroyed by competing interactions and strong quantum fluctuations. In a typical situation, SU(2) symmetry remains intact down to zero temperature so that excitations can be classified by their spin. While magnetic excitations can be studied by techniques such as inelastic neutron scattering (i.e spinons \cite{tianheng,PhysRevB.81.214445}), exotic non-magnetic excitations remain elusive.


Indeed, novel non-magnetic excitations are predicted in generic unconventional magnetic phases. Consider the seminal resonating valence bond (RVB) phase \cite{Anderson1973153}. Extensive investigations of quantum dimer model (QDM) \cite{LMM} revealed that there are two types of dimer liquid in two and three dimensions. The U(1) liquid exists on bipartite lattices in three dimensions. Its low energy excitations are transverse gapless fluctuations of dimer density, or ``photons'' of an emergent U(1) gauge theory \cite{PhysRevB.68.184512}. On the other hand, the Z$_2$ liquid appears in non-bipartite lattices and possesses topological order. The low energy excitations are Z$_2$ vortices, ``visons'' \cite{PhysRevB.40.7133,PhysRevB.62.7850}. While a single vison is a non-local object, excitations of even number of visons correspond to dimer density fluctuations \cite{PhysRevB.70.094430,PhysRevB.71.224109}.

While these results were obtained in the QDM, generic constructions \cite{PhysRevB.72.064413,PhysRevB.80.165131} exist in which QDM's are low energy limits of SU(2) invariant spin models. Such phases and excitations could exist in low energy limits of Heisenberg model thanks to universality. The authors of an extensive DMRG \cite{Yan03062011} study of spin-$1/2$ Heisenberg antiferromagnetic model on the kagome lattice concluded that its ground state is a Z$_2$ spin liquid. Studies of the multi-spin exchange model \cite{PhysRevB.60.1064,PhysRevB.72.045105} on triangular lattice found a gapped spin-liquid phase that looks like a Z$_2$ liquid. Both models are realized in real materials (For reviews, see \cite{LMM} and \cite{nature.464.199}). Observing singlet excitations in these materials would be positive evidence of the existence of Z$_2$ liquid phase in nature.

In this letter, we discuss two general ways to experimentally probe singlet excitations in quantum antiferromagnets. Firstly, singlet excitations mix with optical phonons through spin-lattice coupling. For suitable parameters, this leads to new features in the phonon spectral function which can be detected by neutron or X-ray scattering. Spin-lattice coupling has been an exciting topic throughout the years. Its study was pioneered by the discovery and characterization of spin-Peierls transition \cite{PhysRevB.10.4637,PhysRevLett.35.744,PhysRevB.14.3036,PhysRevB.19.402}. It was realized \cite{PhysRevLett.88.067203,PhysRevB.66.064403} that similar mechanisms can induce long range order in the highly frustrated Heisenberg antiferromagnet on the pyrochlore lattice. Magnetoelastic splitting of degenerate optical phonons was observed in ZnCr$_2$O$_4$ \cite{PhysRevLett.94.137202} and a number of other compounds. The strongest effect (
$10\%$ splitting) has been seen in MnO \cite{rudolf:024421}. Wang and Vishwanath generalized the idea to local phonon  \cite{PhysRevLett.100.077201}. Dynamical effects of phonons were explored \cite{PhysRevB.72.024434} motivated by spin-Peirels compound CuGeO$_3$ \cite{PhysRevLett.70.3651,JPI1996}. These studies demonstrated the importance of both static and dynamical effects of the spin-lattice coupling.

The second way exploits the ability of singlet excitations to couple directly to an electric field. Bulaevskii \emph{et al} \cite{PhysRevB.78.024402} first discovered that some magnetic ground states and excitations of certain Mott insulators have nonzero local electric charge or current. In particular, fluctuations of valence-bond densities induces electric dipoles. The coupling is stronger for weak Mott insulators due to smaller $U/t$ where $t$ is the hopping amplitude of electrons and $U$ is the onsite repulsion. Spin-lattice coupling can lead to the same effect \cite{PhysRevB.78.024402}. Consequently, valence-bond density fluctuations couple to electromagnetic radiations directly. Conventional spectroscopic techniques may be used to directly detect non-magnetic excitations.

The rest of the paper is organized as follows. First, we discuss the mixing between optical phonons and singlet excitations. After introducing a general formulation, we explore its consequences for both U(1) and Z$_2$ liquids. We then discuss possible charge signature of singlet excitations in the context of spin-$1/2$ Heisenberg antiferromagnetic model on kagome. Finally, we conclude our paper by discussing possible discoveries of singlet excitations in real materials.

\emph{Optical phonon:}
Let us motivate the first mechanism in the simplest context. Consider spins interacting via Heisenberg exchange whose strength depends on the distance between the spins \cite{PhysRevLett.88.067203,PhysRevB.66.064403,PhysRevLett.94.137202}:
\begin{equation}\label{eqn:singlebond}
    J(R+u)\mathbf{S}_i\cdot\mathbf{S}_j\approx J(R)\mathbf{S}_i\cdot\mathbf{S}_j+\frac{\partial J}{\partial r}|_{r=R}(\mathbf{S}_i\cdot\mathbf{S}_j)u.
\end{equation}
$R$ is the equilibrium distance between the two spins and $u$ is the elongation of the bond. $u$ couples linearly with bond operator $\mathbf{S}_i\cdot\mathbf{S}_j$ which measures the singlet density on bond $\langle ij\rangle$. The coupling mixes the singlet excitations and phonons. The phonon spectral function will acquire features of the singlet excitations.

To elaborate on this idea, we consider following Hamiltonian on a general lattice:
\begin{equation}\label{eqn:hami}
    H=\sum_{\mathbf{r}}\left(\frac{1}{2}\dot{\mathbf{u}}(\mathbf{r})^2+\frac{1}{2}\omega_0^2\mathbf{u}(\mathbf{r})^2+f\mathbf{u}(\mathbf{r})\cdot \mathbf{V}(\mathbf{r})\right)+H_{s}.
\end{equation}
$H_{s}$ is the spin hamiltonian and $\mathbf{u}(\mathbf{r})$ is the displacement of ion at site $\mathbf{r}$. We adopt the Einstein phonon model with mass of the ion assumed to be $1$ for simplicity. Based on the model in equation \ref{eqn:singlebond}, $\mathbf{V}(\mathbf{r})$ field is defined as follows \cite{PhysRevLett.100.077201}:
\begin{equation}\label{eqn:vector}
    \mathbf{V}(\mathbf{r})=\frac{1}{f}\sum_{\mathbf{r}^\prime\in\{\mathbf{r}\}}\hat{e}_{\mathbf{r}\mathbf{r}^\prime}(\hat{e}_{\mathbf{r}\mathbf{r}^\prime}\cdot\nabla_{\mathbf{r}} J(\mathbf{r}-\mathbf{r}^\prime))\mathbf{S}_{\mathbf{r}}\cdot\mathbf{S}_{\mathbf{r}^\prime}.
\end{equation}
$\{\mathbf{r}\}$ is the set of neighbors of site $\mathbf{r}$ and $\hat{e}_{\mathbf{r}\mathbf{r}^\prime}\equiv (\mathbf{r}-\mathbf{r}^\prime)/|\mathbf{r}-\mathbf{r}^\prime|$. $f=\sum_{\mathbf{r}^\prime\in\{\mathbf{r}\}}\hat{e}_{\mathbf{r}\mathbf{r}^\prime}\cdot\nabla_{\mathbf{r}} J_{\mathbf{r}-\mathbf{r}^\prime}$ is the spin-lattice coupling.


For small ion displacements, the full phonon Green's function in the random phase approximation (RPA) is:
\begin{equation}\label{eqn:fullphonon}
    \tilde{G}^{-1}_{\alpha\beta}(\mathbf{r}_1,t_1;\mathbf{r}_2,t_2)=G^{-1}_{\alpha\beta}(\mathbf{r}_1,t_1;\mathbf{r}_2,t_2)-f^2\chi_{\alpha\beta}(\mathbf{r}_1,t_1;\mathbf{r}_2,t_2)
\end{equation}
where $\tilde{G}$ is the full phonon Green's function, $G$ is the bare one and $\chi$ is the time-ordered bond-bond correlation function:
\begin{equation}\label{eqn:dynsus}
    \chi_{\alpha\beta}(\mathbf{r}_1,t_1;\mathbf{r}_2,t_2)\equiv \langle \mathrm{T}\{V_{\alpha}(\mathbf{r}_1,t_1)V_{\beta}(\mathbf{r}_2,t_2)\}\rangle
\end{equation}
In the Fourier space, relation \ref{eqn:fullphonon} is written as:
\begin{equation}\label{eqn:fullphononmomentum}
    \tilde{G}^{-1}_{\alpha\beta}(\mathbf{k},\omega)=G^{-1}_{\alpha\beta}(\mathbf{k},\omega)-f^2\chi_{\alpha\beta}(\mathbf{k},\omega).
\end{equation}

We explore the consequences of eqn \ref{eqn:fullphonon} and \ref{eqn:fullphononmomentum} in two examples in QDM's. QDM's \cite{LMM} are low-energy effective models describing the dynamics of nearest neighbor singlets. They provide generic playgrounds \emph{independent} of the underlying spin models.

The first example is U(1) RVB liquid phase on cubic lattice \cite{PhysRevLett.91.167004,PhysRevB.68.184512}. The QDM on cubic lattice has two phases \cite{PhysRevB.68.184512}: the staggered valence bond crystal phase and the U(1) liquid phase.  The low energy physics of the liquid phase is described by the following Hamiltonian in the continuum limit \cite{PhysRevB.68.184512}:
\begin{equation}\label{eqn:u1hami}
    H_s=\int d^3r\left(\frac{1}{2}\mathbf{E}^2+\frac{1}{2}\rho_2\mathbf{B}^2+\rho_4(\nabla\times\mathbf{B})^2\right).
\end{equation}
In the Coulomb gauge $A_0=0$, $\nabla\cdot\mathbf{A}=0$, the electric and magnetic fields are expressed as $\mathbf{E}=\partial_t\mathbf{A}$ and $\mathbf{B}=\nabla\times\mathbf{A}$. On the lattice, the magnetic field $\mathbf{B}$ is defined on the bonds \cite{PhysRevLett.91.167004}:
\begin{equation}\label{eqn:Bfield}
    B_{\alpha}(\mathbf{r})=e^{i\mathbf{Q}\cdot\mathbf{r}}\left(n_{\alpha}(\mathbf{r})-\frac{1}{z}\right)
\end{equation}
where $\mathbf{Q}=(\pi,\pi,\pi)$ and $n_{\alpha}(\mathbf{r})$ is the number of dimers on the bond connecting $\mathbf{r}$ and $\mathbf{r}+\hat{\alpha}$ ($\alpha=x,y,z$). $z=6$ is the coordination number of cubic lattice.
\begin{figure}
  \includegraphics[width=0.7\columnwidth]{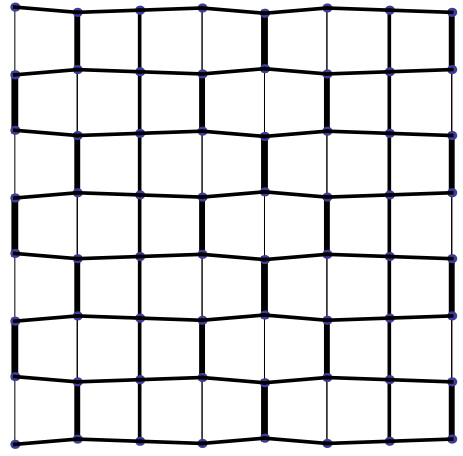}\\
  \caption{The new phonon mode around $\mathbf{Q}$ is illustrated on a two dimensional slice of cubic lattice. The mode is induced by the ``photon'', transverse fluctuation of dimer densities. Bond thicknesses reflect dimer densities. }\label{fig:squarephonon}
\end{figure}

To include the spin-lattice coupling, we write the bond operator $\mathbf{S}(\mathbf{r})\cdot\mathbf{S}(\mathbf{r}+\hat{\alpha})$ in terms of the gauge field. The bond operator amounts to two operations on a general dimer covering. Its diagonal part counts the number of dimer on the bond while the off-diagonal part flips the dimers around plaquette to which the bond belongs. The plaquette-flipping operator translates into $\mathbf{E}^2$ \cite{PhysRevB.68.184512}. As a result, the off-diagonal term is irrelevant in the renormalization group sense. The spin-lattice Hamiltonian translates into the following compact form in the continuum limit:
\begin{equation}\label{eqn:spinlattice}
    H_{sp}=\int d^3r f_1\mathbf{B}\cdot\mathbf{\tilde{u}}
\end{equation}
where $\mathbf{\tilde{u}}\equiv e^{i\mathbf{Q}\cdot\mathbf{r}}\mathbf{u}$ and $f_1$ is the spin-lattice coupling. The Hamiltonian for phonon is:
\begin{equation}\label{eqn:phonon}
    H_p=\int d^3r \left(\frac{1}{2}(\partial_t\mathbf{\tilde{u}})^2+\frac{1}{2}\omega_0^2\mathbf{\tilde{u}}^2\right).
\end{equation}
The total Hamiltonian is $H=H_p+H_{sp}+H_{s}$.

We focus on the phonon spectrum in U(1) liquid phase where $\rho_4$ term can be neglected. Applying equations \ref{eqn:fullphonon} and \ref{eqn:fullphononmomentum}, phonon develops two transverse sound modes around momentum $\mathbf{Q}$. These modes manifest themselves as new low-energy poles in the phonon spectral function.  For momentum $\mathbf{Q}+\mathbf{k}$ ($k\ll 1$), the energy is approximately:
\begin{equation}\label{eqn:dispersion}
    \omega(\mathbf{k})\approx\sqrt{\rho_2-\frac{f_1^2}{\omega_0^2}}k.
\end{equation}
The spectrum weight of the modes is proportional to $f_1^2k^2/(\omega_0^2-\rho_2k^2)^2$. In contrast, the spectrum of the longitudinal phonon remains unchanged. This is a reflection of the transverse nature of gauge fluctuations. Such sound modes generally exist in QDM's on other three-dimensional bipartite lattices.

If the spin-lattice coupling is so large that $f_1^2>\rho_2\omega_0^2$, the lattice distorts and the magnetic fluxes condenses. This is the analog of spin-Peierls transition. We will discuss it in a future publication.

\begin{figure}
  \centering
  \includegraphics[width=0.8\columnwidth]{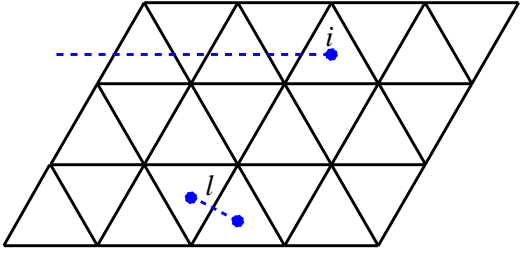}\\
  \caption{One-vison and two-vison excitations. A vison at $i$ is created by operators on all the bonds to the left of $i$. A two-vison excitation is related to the dimer density on the bond $l$ between them.}\label{fig:vison}
\end{figure}

The second generic dimer liquid phase is the Z$_2$ liquid phase on two and three dimensional non-bipartite lattices \cite{LMM}. Such a state preserves all lattice symmetries and has a gap to all excitations. The system possesses topological order. Consider a Z$_2$ liquid on a cylinder, the state belongs to the even or odd topological sectors if a cut around the cylinder crosses even or odd number of dimers. The low energy excitations are visons \cite{PhysRevB.40.7133,PhysRevB.62.7850}, Z$_2$ vortices residing on the sites of the dual lattices. At the Rokhsar-Kivelson (RK) \cite{PhysRevLett.61.2376} point, the ground state is an equal amplitude combination of all possible dimer state. To obtain a vison at site $\mathbf{\tilde{r}}$, we make a cut \cite{LMM} from the site to the lattice boundary (Fig \ref{fig:vison}). A state with one vison is a combination of all possible dimer coverings. However, the amplitude of a dimer covering is negative if the cut crosses an odd number of dimers. Visons are nonlocal objects. A local operator such as $\mathbf{S}_{\mathbf{r}}\cdot\mathbf{S}_{\mathbf{r}+\hat{l}}$ involves even number of visons.

We consider two cases. In the first scenario, two-vison excitations have definite energy and momentum. This makes sense if visons form bound pairs. Such a bound state has been observed \cite{Lauchli2008} in numerical studies of QDM on triangular lattice. It is also possible that the vison dispersion is almost flat \cite{PhysRevB.68.214415,PhysRevB.73.245103,PhysRevB.84.094419} so that the two-vison continuum is very narrow. In the former case, the two-vison excitation induces a new pole in the phonon spectrum in the same fashion as the U(1) case.  For the latter, a very narrow continuum will be present in the low-energy spectral function of phonons. The spectral weight of either the pole or the narrow continuum is determined by $\omega_v^4/\omega_0^4$ where $\omega_v$ is the energy of the two-vison excitation.

If the two-vison excitations form a broad continuum, the vison has a large bandwidth with $N$ minima in the first Brillouin zone. We introduce coarse-grained field operators around these minima:
\begin{equation}\label{eqn:fieldoperator}
    \phi_{\alpha}(\mathbf{r})=\frac{1}{\sqrt{N}}\sum_{\mathbf{q}}\frac{1}{\sqrt{2\omega_\mathbf{q}}}a_{\alpha,\mathbf{q}} \mathrm{e}^{-\frac{q}{\Lambda}}\mathrm{e}^{i\mathbf{q}\cdot\mathbf{r}}
\end{equation}
where $\alpha=1\ldots N$ and $\Lambda$ is the momentum cut-off set by the lattice scale. In terms of the field operators, the vison-phonon coupling can be written in the following form:
\begin{eqnarray}\label{eqn:lowenergycoupling}
    \nonumber H=&\int d^dr \sum_{l=x,y} \sum_{\alpha\ge \beta}u_{l}(\mathbf{r})\left(g_{\alpha\beta}^{(l)}\mathrm{e}^{i(\mathbf{k}_\alpha+\mathbf{k}_\beta)}\phi_\alpha(\mathbf{r})\phi_\beta(\mathbf{r})\right.\\ &\left.+f_{\alpha\beta}^{(l)}\mathrm{e}^{-i(\mathbf{k}_\alpha-\mathbf{k}_\beta)}\phi^\ast_\alpha(\mathbf{r})\phi_\beta(\mathbf{r})+\mathrm{h.c}\right).
\end{eqnarray}
$g_{\alpha\beta}^{(l)}$ and $f_{\alpha\beta}^{(l)}$ can be derived from a microscopic model. They can also be fixed up to an overall constant by symmetries. The energy of vison fields is $\omega_\mathbf{q}=\sqrt{m_\alpha^2+v_\alpha^2q^2}$. For simplicity, we assume $m_\alpha=m$ and $v_\alpha=v$. Since visons are gapped, only $\phi_{\alpha}\phi_\beta$ and its complex conjugate contribute to $\chi$.

As $\omega$ approaches the two-vison continuum from below, $\chi(\omega,\mathbf{k}_\alpha+\mathbf{k}_\beta)$ diverges logarithmically at $\omega=2m$. This leads to appearance of a new pole in phonon Green's function \ref{eqn:fullphonon}. The residue of the new mode increases as the pole moves away from $2m$. For appropriate parameters, such modes can be observed.

As $m$ decreases, the new phonon modes are pushed toward zero energy. At a critical $m$, one (or several) of them condenses. The system becomes a valence bond crystal. This is the analog of Spin-Peierls transition in Z$_2$ liquid. We will discuss it in detail in a future work.

\begin{figure}
  \includegraphics[width=0.8\columnwidth]{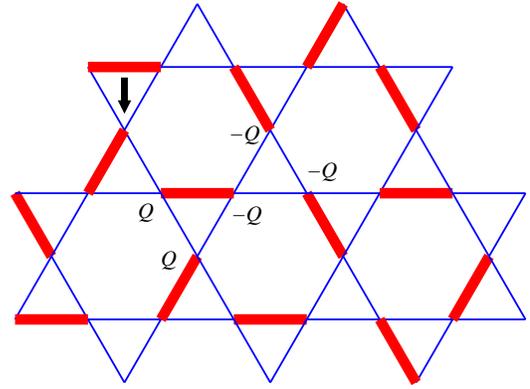}\\
  \caption{A dimer-covering state of the kagome lattice. A vacuum triangle carries an electric dipole illustrated by the black arrow assuming $Q>0$. Vertices around the defect triangle carry charge $-Q$ while other vertices carry $Q$. }\label{fig:charge}
\end{figure}

\emph{Charge signature:}
On non-bipartite lattices, singlet density fluctuations generate electric dipoles \cite{PhysRevB.78.024402}.  We consider three spins on a equal lateral triangle interacting antiferromagnetically. The exchange energy is minimized by combining two of them into a singlet. The induced electric dipole lies in the plane of the triangle normally to the singlet bond \cite{PhysRevB.78.024402} (Figure \ref{fig:charge}). In other words, charge $2Q$ accumulates on the free spin while the two spins forming the singlet carry charge $-Q$ each. If the dipole is induced by higher-order perturbations in a weak Mott insulator, $Q>0$ is proportional to $(t/U)^3$. The sign and the magnitude of spin-lattice coupling determines $Q$ when the dipole is generated by magnetostriction.

Consider spin-$1/2$ Heisenberg antiferromagnetic model on the kagome lattice (Figure \ref{fig:charge}), a network of corner sharing triangles. The low energy states are dimer-covering states with maximum number of nearest neighbor singlets. A quarter of triangles, the so-called ``defect triangles'', lack singlets\cite{PhysRevB.48.13647,PhysRevLett.103.187203}. The uneven distribution of singlet densitiestranslates into inhomogeneous charge densities. A simple counting shows that the vertices of defect triangles each carry $-Q$ while all other vertices have $Q$.

An applied AC electric field can be used to induce the motion of defect triangles. For typical sample size ($L=1$ mm) and singlet excitation energy ($\omega=0.1$ meV), the applied field is approximately uniform, $q=\omega/c\gg 1/L$. The scattering of the applied field will provide information about the singlet spectrum of the system at $\mathbf{q}=0$. Suitable techniques include, for example, Terahertz spectroscopy.

\emph{Discussion:}
In this letter we described how exotic singlet excitations can be detected, at least in principle, by  existing spectroscopic methods. While focusing on specific models, we stress that the two mechanisms described are \emph{model-independent} as long as the global SU(2) symmetry is intact.

Currently two classes of materials could host the Z$_2$ liquid state. The first set of materials including ZnCu$_3$(OH)$_6$Cl$_2$ (herbertsmithite), Cu$_3$V$_2$O$_7$(OH)$_2\cdot$2H$_2$O (volborithite) and BaCu$_3$V$_2$O$_8$(OH)$_2$ (vesignieite) (See \cite{LMM} and \cite{nature.464.199} for reviews) realize the $S=1/2$ Heisenberg antiferromagnetic model on the kagome lattice. Yan \textit{et al} \cite{Yan03062011} presented some evidence that the ground state of the model is a Z$_2$ spin liquid. The vison spectrum were studied by several groups \cite{PhysRevB.68.214415,PhysRevB.73.245103,PhysRevB.84.094419}. They identified the location of low energy singlet excitations in the reciprocal space. Recently, single-crystal sample of herbertsmithite was synthesized \cite{PhysRevB.83.100402}. Measuring the spectral function of phonons in the magnetic energy range (up to a few meV) could reveal novel singlet excitations. It would also be interesting to measure its spectrum at $\mathbf{q}=0$ using conventional spectroscopies.

The second class of materials including $\kappa$-(BEDT-TTF)$_2$Cu$_2$(CN)$_3$ and EtMe$_3$Sb[Pd(dmit)$_2$]$_2$ (See \cite{nature.464.199} for a review) realizes the multi-spin exchange model on triangular lattice. Studies \cite{PhysRevB.60.1064,PhysRevB.72.045105} show that the ground state of the model can be a gapless or a gapped spin liquid for different parameters. The gapped liquid phase resembles the Z$_2$ liquid phase with a large number of singlet excitations within the spin gap \cite{PhysRevB.60.1064}. While $\kappa$-(BEDT-TTF)$_2$Cu$_2$(CN)$_3$ \cite{PhysRevLett.91.107001} seems to host the gapless liquid phase, the flexibility of the material family $\kappa$-(BEDT-TTF)$_2$X raises the hope that the gapped liquid phase is the ground state for some other member whose singlet excitations can be observed by studying the phonon spectrum.

Beyond quantum magnetism, singlet excitations are believed to be important for other strongly correlated systems such as high temperature superconductors \cite{RevModPhys.78.17}. We speculate similar couplings between optical phonons and singlet excitations also exist. It would be very interesting to search for the trace of singlet excitations in phonon spectrum in these systems.

I acknowledge helpful discussions with Oleg Tchernyshyov, Michel Gingras, Collin Broholm and Natalia Drichko. Oleg Tchernyshyov is specially thanked for careful read and thorough critiques of an early draft. The work is supported by the U.S. Department of Energy, Office of Basic Energy Sciences, Division of Materials Sciences and Engineering under Award No. DE-FG02-08ER46544 and NSERC of Canada.

\bibliography{sl}
\end{document}